\documentclass[aps,prl,reprint,amsmath,amssymb]{revtex4-1}

\usepackage{graphicx}
\usepackage{bm}

\newcommand{\gd}{g^{(2)}}
\newcommand{\pit}{2\pi \times}
\newcommand{\nth}{n_{\rm th}}
\newcommand{\ntot}{n_{\rm tot}}

\begin{document}

\title{Observation of the Unconventional Photon Blockade in the Microwave Domain}

\author{Cyril Vaneph}
\author{Alexis Morvan}
\author{Gianluca Aiello}
\author{Mathieu F\'echant}
\author{Marco Aprili}
\author{Julien Gabelli}
\author{J\'er\^ome Est\`eve}
\affiliation{Laboratoire de Physique des Solides, CNRS, Universit\'e Paris-Sud, Universit\'e Paris-Saclay, Orsay, France}
\date{\today}

\begin{abstract}
We have observed the unconventional photon blockade effect for microwave photons using two coupled superconducting resonators. As opposed to the conventional blockade, only weakly nonlinear resonators are required. The blockade is revealed through measurements of the second order correlation function $\gd(t)$ of the microwave field inside one of the two resonators. The lowest measured value of $\gd(0)$ is 0.4 for a resonator population of approximately $10^{-2}$ photons. The time evolution of $\gd(t)$ exhibits an oscillatory behavior, which is characteristic of the unconventional photon blockade. 
\end{abstract}

\maketitle
Photon blockade is observed when a single two-level emitter, such as an atom \cite{Birnbaum:2005cj}, a quantum dot \cite{Faraon:2008da}, or a superconducting qubit \cite{Lang:2011iq,Hoffman:2011fz} is strongly coupled to a cavity, thus limiting the occupation of the cavity mode to zero or one photon. The second order correlation function $\gd(t)$ of the light leaking out of the cavity shows a dip at short time with $\gd(0)<1$, a signature of nonclassical fluctuations corresponding to antibunched photons. The same effect is predicted for a nonlinear Kerr cavity when the Kerr nonlinearity $U$ is much larger than the cavity linewidth $\kappa$ \cite{Imamoglu:1997gp}. In 2010, Liew and Savona discovered that this constraint can be relaxed by considering two coupled cavities instead of one \cite{Liew:2010kg}. They found that perfect blockade $\gd(0)=0$ can be achieved even for a vanishingly small ratio $U/\kappa$ and named the effect "unconventional photon blockade" (UPB). The UPB was later interpreted as an interference between the two possible paths from the one to the two photon state \cite{Bamba:2011gg} or as the fact that the cavity state is a displaced squeezed state \cite{Lemonde:2014wj}. Such states are known to exhibit antibunching for well-chosen displacement and squeezing parameters \cite{Stoler:1974bz,Mahran:1986cq,Lu:2001de,Grosse:2007fy}. Reaching the strong coupling regime between a cavity and an emitter, or a large $U/\kappa$ in a Kerr cavity, remains highly challenging, especially in the optical domain. Therefore the UPB has attracted considerable attention \cite{Flayac:2017kt} by opening new possibilities to obtain sources of nonclassical light using readily available nonlinear cavities forming a photonic molecule \cite{Galbiati:2012js,Adiyatullin:2017ww}.

\begin{figure}[ht!]
\includegraphics[scale=0.94]{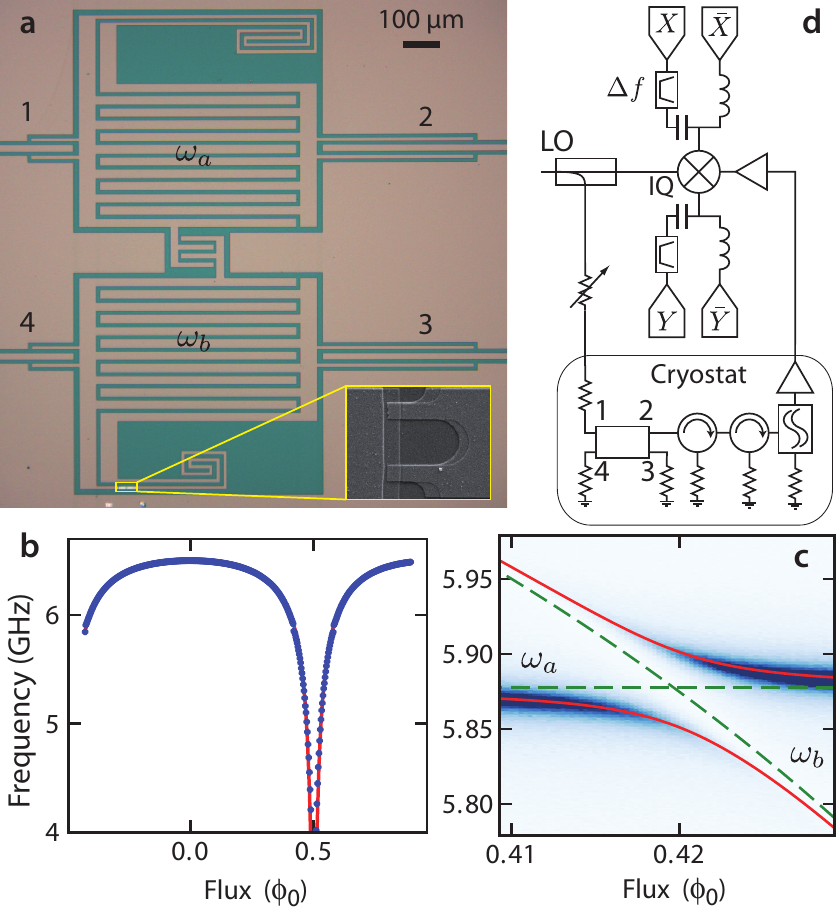}
\caption{a) Microscope image of the two coupled Nb resonators used to observe the UPB. The bottom resonator is frequency tunable and slightly nonlinear due to the presence of a SQUID in the inductive arm. The SQUID consists of two Al/AlOx/Al Josephson junctions with a surface of 1\,${\rm \mu m}^2$ each. The blockade occurs in the top resonator, which is linear. The interdigited capacitance in the center couples the two resonators. The numbers label the ports connected to the sample. b) Evolution of the resonance frequency $\omega_b$ as a function of the SQUID flux. c) Evolution of $|S_{12}|^2$ as a function of frequency and SQUID flux. d) Microwave setup used to measure the UPB. Ports 3 and 4 are terminated by 50\,$\Omega$ loads anchored at 10\,mK. The LO signal is attenuated and pumps the system through the port 1. The signal of interest exits through port 2 and goes through two circulators and a diplexer before reaching the amplifier. After further amplification outside the cryostat, the signal is mixed with the LO and the resulting dc and ac components of each quadrature are filtered and digitized.}
\end{figure}

Here, we report on the observation of the UPB for microwave photons in a superconducting circuit consisting of two coupled resonators, one being linear and one weakly nonlinear \cite{Eichler:2014dk}. We measure the moments of the two quadratures of the field inside the linear resonator using a linear amplifier \cite{daSilva:2010io,Eichler:2012cs}. The determination of $\gd(0)$ for an arbitrary field requires measuring the moments of the two quadratures up to the fourth order. But in the case of the UPB, the state of the field is expected to be a displaced squeezed Gaussian state, therefore the value of $\gd(0)$ can be accurately obtained from the measurement of the first and second order moments only. This greatly reduces the experimental acquisition time and allows us to perform an exhaustive study of the blockade phenomenon as a function of various experimental parameters. In particular, we have searched for the optimal $\gd(0)$ as a function of the resonator population. We also measure $\gd(t)$ and observe oscillations that are characteristic of the UPB. Finally, we confirm the validity of the Gaussian assumption through measurements of the moments up to the fourth order. 

Figure~1a shows a microscope image of the sample. Two resonators made of niobium and consisting of an inductance in series with a capacitance are coupled through a capacitance. The inductive part of the bottom resonator includes a SQUID that introduces a Kerr nonlinearity. Both resonators are coupled to two coplanar waveguides (CPW) that allow us to pump and probe the resonator fields. The effective Hamiltonian of the circuit is  
\begin{equation}
H/\hbar = \omega_a a^\dagger a + \omega_b b^\dagger b + J (a^\dagger b + b^\dagger a) - U b^\dagger b^\dagger bb,
\label{eq.H}
\end{equation} 
where $\omega_a$ is the resonance frequency of the top resonator, $\omega_b$ is the resonance of the bottom resonator, which depends on the SQUID flux, $J$ the coupling and $U$ the Kerr nonlinearity. As shown in \cite{Bamba:2011gg}, this Hamiltonian leads to perfect blockade under the condition $\omega_a=\omega_b$ and $U=2\kappa^3/(3\sqrt{3}J^2)$, where $\kappa$ is the loss rate of the resonators. The sample was designed to fulfill this condition with $J=\pit 25$\,MHz, $\kappa=\pit 8$\,MHz and $U=\pit 0.3$\,MHz. 

To check these values for our sample, we first measure the evolution of $\omega_b$ with the SQUID flux as shown in figure 1b. We assume that the bottom resonator can be modeled by a lumped element circuit formed by the association in series of a capacitor $C$, an inductance $L$ and the SQUID inductance $L_s$, which varies with the applied flux $\phi$ as $L_s = L_{s0}/|\cos(\pi \phi/\phi_0)|$. From the red fit, we obtain $L=1.09$\,nH and $L_{s0}=81$\,pH. When $\omega_b\approx \omega_a=\pit 5.878$\,GHz, we obtain $L_s=337$\,pH, from which we deduce the Kerr nonlinearity $U=\pi p^3/(2R_KC)=\pit 0.25$\,MHz, where $R_K=h/e^2\approx 25.8\,{\rm k}\Omega$ and $p=L_s/(L+L_s)$ \cite{Ong:2011cf}. Figure 1c shows a measurement of the top resonator transmission when $\omega_b$ crosses $\omega_a$. By fitting the observed avoided level crossing, we obtain $J=\pit 25.1$\,MHz. Finally, we have measured the linewidths (fwhm) of each resonator and obtained $\kappa_a = \pit 10.35$\,MHz for the top resonator and $\kappa_b = \pit 7$\,MHz for the bottom resonator. 

\begin{figure}[t!]
\includegraphics[scale=0.95]{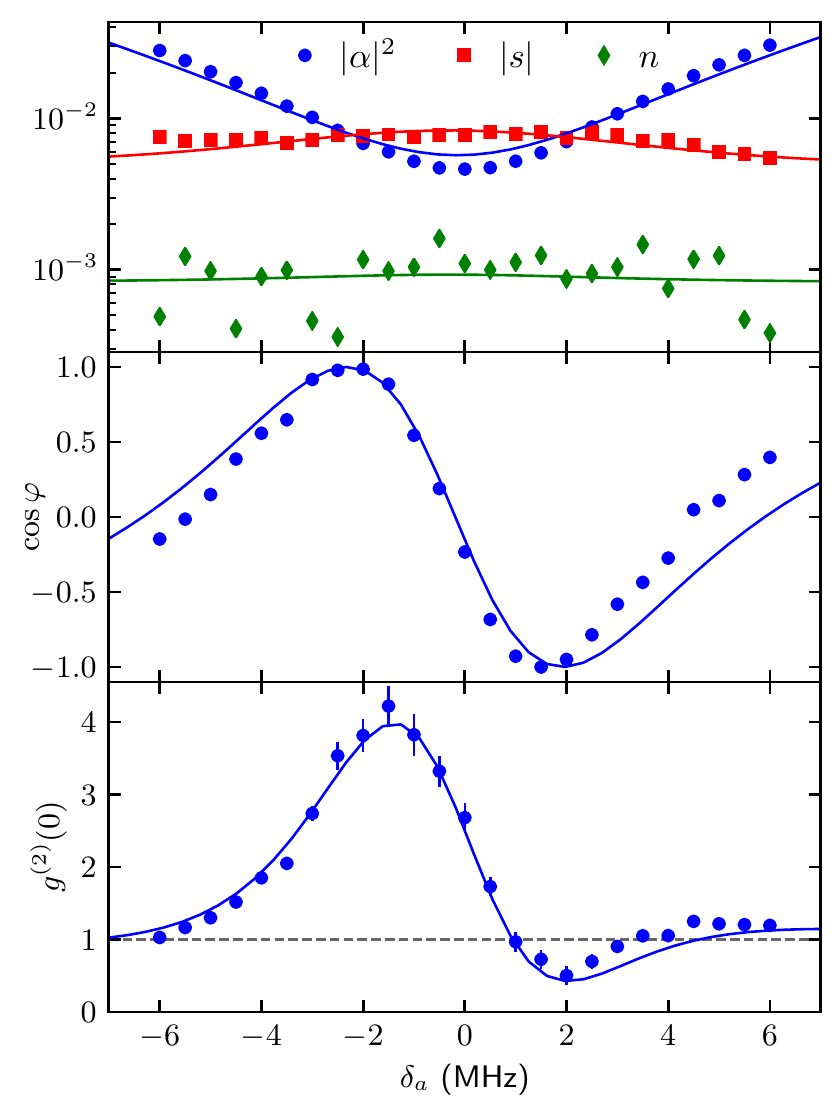}
\caption{Evolution of $\gd(0)$ with the pump detuning. The two uppermost plots show the evolution of the Gaussian parameters characterizing the Gaussian state (see text) in the resonator as a function of $\delta_a$. Markers correspond to experimental data points and solid lines to the solution of the master equation \cite{Note1}. From these  quantities, we compute $\gd(0)$ using equation (\ref{eq.g2Gaussian}). The error bars correspond to statistical $1\sigma$ errors. The nonlinear resonator is tuned to $\omega_b\approx \omega_a$ within a few MHz and the incident pump power on the sample is -107\,dBm.}
\end{figure}

In order to measure the UPB, we make the assumption that the state in the resonator is Gaussian. This assumption is well verified in numerical simulations of the master equation describing our system \footnote{See Supplemental Material for details.} in accordance with the predictions of \cite{Lemonde:2014wj}. Therefore, the quantum state of the resonator field $a$ is characterized by the displacement $\alpha =\langle a \rangle$ and by the Gaussian noise ellipse around the mean displacement. Defining the operator $d=a-\alpha$, the fluctuations of $d$ are Gaussian and are the one of a squeezed thermal state, which can be parametrized by the real number $n=\langle d^\dagger d\rangle$ and the complex number $s=\langle d d \rangle$. In the case of our experiment, because $s$ remains small, $n$ is the population of the thermal state. With these definitions, the second order correlation function at zero time is
\begin{equation}
\gd(0) = 1+\frac{2|\alpha|^2(n+|s|\cos \varphi)+|s|^2+n^2}{(|\alpha|^2+n)^2},
\label{eq.g2Gaussian}
\end{equation}
where $\varphi$ is the complex argument of $s/\alpha^2$ \cite{Stoler:1974bz}. This formula shows that a finite amount of squeezing is necessary to have $\gd(0)<1$. In the limit of a squeezed state with minimal uncertainty $n=0$ and supposing $|s|=|\alpha|^2$, one obtains $\gd(0) = 2+2\cos \varphi$ showing that $\gd(0)$ oscillates with $\varphi$ between 0 and 4. Perfect antibunching is obtained when the state simultaneously fulfills the two conditions $|s|=|\alpha|^2$ and $\cos \varphi = -1$. Experimentally, one has to tune the pump and the nonlinear resonator frequencies to meet these conditions.  

The measurement of $\alpha$, $n$ and $s$ is performed by amplifying the field leaving the top resonator with a cryogenic amplifier and by measuring the two quadratures of the amplified field as shown in figure 1d. We suppose that the field at the input of the IQ mixer is proportional to $a+h^\dagger$, where $h$ is a Gaussian field whose fluctuations are dominated by the intrinsic noise of the amplifier \cite{daSilva:2010io,Eichler:2012cs}. At the output of the mixer, we separate the ac and the dc components of each quadrature. The dc components $\bar{X}$, $\bar{Y}$ measure $\alpha$ while the ac components $X$, $Y$ are the quadratures of the field $d+h^\dagger$ at the pumping frequency. As shown in Ref. \cite{Lemonde:2014wj}, the noise spectrum of $d$ consists of two peaks centered approximately at $\pm J$ with a linewidth $\kappa$. We therefore filter the AC components with a bandpass filter of bandwidth $\Delta f=24$\,MHz centered at 22.5\,MHz.

\begin{figure}[t]
\includegraphics[scale=1]{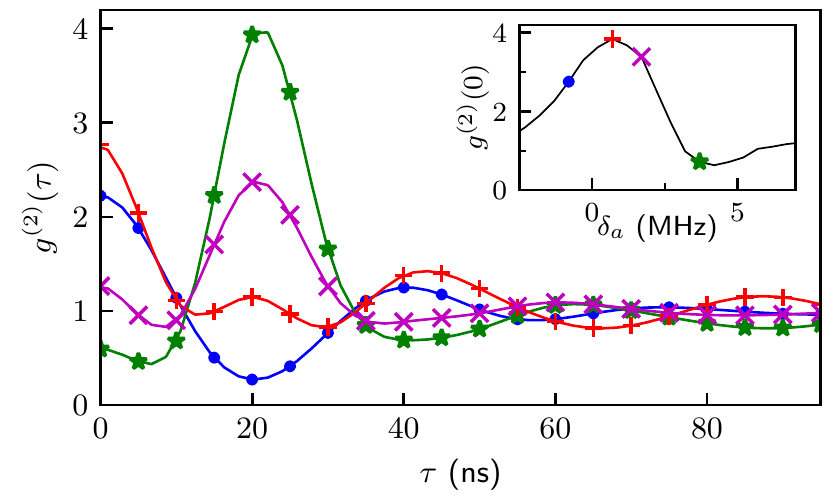}
\caption{Measured time evolution of $\gd(\tau)$ for four different pump detunings. Solid lines interpolate the experimental data points. The bottom resonator is tuned to $\omega_b\approx\omega_a$ and the incident pump power is -101\,dBm. Each curve corresponds to a different detuning $\delta_a$. The inset shows the evolution of $\gd(0)$ as a function of $\delta_a$, the four detunings corresponding to the curves in the main plot are identified by colored points. The oscillation of $\gd(\tau)$ with time is characteristic of the UPB. Depending on the initial phase of the oscillation, the state violates none, one or two of the inequalities $\gd(0)\ge1$, $\gd(0)\ge\gd(\tau)$ that can be derived for a classical field. }
\end{figure}

The population of $h$ is $n_h= 2 k_B T_{\rm ampl} \Delta f/(G_{\rm att} \gamma_2)$, where $T_{\rm ampl}=2$\,K is the amplifier noise temperature,  $G_{\rm att}=-3$\,dB is the attenuation between the sample and the amplifier, and $\gamma_{2}=\pit 8.6$\,MHz is the simulated loss rate from the mode $a$ to the measurement port \cite{Note1}. These values lead to $n_h=12.5$, which must be compared to the expected values $|s|\approx 10^{-2}$ and $n\approx 10^{-3}$. In order to extract this small signal, we alternately acquire data turning the pump on and off and repeat this cycle many times. The period of the cycle is kept below 1\,s to avoid any influence of a drift of the amplifier noise or gain. The expression of $\alpha$, $n$ and $s$ as a function of the measured moments are 
\begin{subequations}
\label{eq.Gaussianp}
\begin{eqnarray}
\alpha & = & \frac{\langle \bar{X} \rangle_1 + i  \langle \bar{Y} \rangle_1}{\sqrt{2}}  \\
n & = & \frac{\langle X^2\rangle_1 -\langle X^2\rangle_0 + \langle Y^2\rangle_1 -\langle Y^2\rangle_0}{2} + \nth \\ 
s & = & \frac{\langle X^2\rangle_1 -\langle X^2\rangle_0 - \langle Y^2\rangle_1 +\langle Y^2\rangle_0}{2} + i\langle XY \rangle_1
\end{eqnarray}
\end{subequations}
where $\langle \cdot \rangle_1$ ($\langle \cdot \rangle_0$) corresponds to averaged data when the pump is on (off). The data are rescaled to correct for the imperfections of the IQ mixer such that $\langle X^2\rangle_0=\langle Y^2\rangle_0= n_h$ and $\langle XY \rangle_0=0$ \cite{Note1}. By construction, the measurement of $n$ is only sensitive to a relative change of the fluctuations of the resonator field. We therefore have to make an assumption for the occupation of the measured mode when the pump is off. We suppose that the population is thermal with a mean occupation $\nth$ that we calculate by estimating the incident thermal radiation on both resonators and solving the master equation \cite{Note1}. We obtain $\nth=7.8 \times 10^{-4}$, which corresponds to a temperature of 39.4\,mK.

Figure 2 shows the results of the measurement of the Gaussian parameters $\alpha$, $n$ and $s$ when $\omega_a \approx \omega_b$ as a function of the detuning $\delta_a=\omega_p-\omega_a$ where $\omega_p$ is the pump frequency. The amplitude of the field $|\alpha|^2$ passes by a minimum when the detuning increases. Around this minimum, $|\alpha|^2$ is on the order of $|s|$ and $\gd(0)$ deviates significantly from one. The angle $\varphi$ determines the sign of the deviation and its evolution explains the oscillation of $\gd(0)$ around the resonance. The amount of squeezing is small and the Wigner distribution of the state is almost an isotropic Gaussian function. But because the displacement is also small, the squeezing is sufficient to make the overlap between the Wigner distribution of the state and the one of the two-photon Fock state smaller than for a coherent state. This happens when the small axis of the squeezing ellipse is aligned with the direction of the displacement $\alpha$ in the $XY$ plane, resulting in $\gd(0)<1$. Simulations of the master equation using the measured values for $U$,$J$,$\kappa_a$ and $\kappa_b$ well reproduce the observed evolution. The only adjustable parameter in the simulation is the pump intensity that we adjust to reproduce the observed displacement $|\alpha|^2$.

The Gaussian assumption can be extended to the measurement of $\gd(\tau)$ by introducing the time-dependent quantities $n(\tau)$ and $s(\tau)$. They are defined from the measured time-dependent correlation as in Eq. (\ref{eq.Gaussianp}) with the transformation $\langle X^2\rangle \rightarrow \langle X(t)X(t+\tau) \rangle$, $\langle Y^2\rangle \rightarrow \langle Y(t)Y(t+\tau) \rangle$ and $\langle XY \rangle \rightarrow \langle X(t)Y(t+\tau)\rangle$. The results are plotted in figure 3 for four different pump frequencies. Because the squeezing results from the interference between the two components of the noise spectrum at $+J$ and $-J$, the angle $\varphi$ oscillates in time with a period $2\pi/J=40$\,ns resulting in an oscillation of $\gd(\tau)$ that is characteristic of the UPB \cite{Liew:2010kg}.

\onecolumngrid

\begin{figure}[t]
\includegraphics[scale=1]{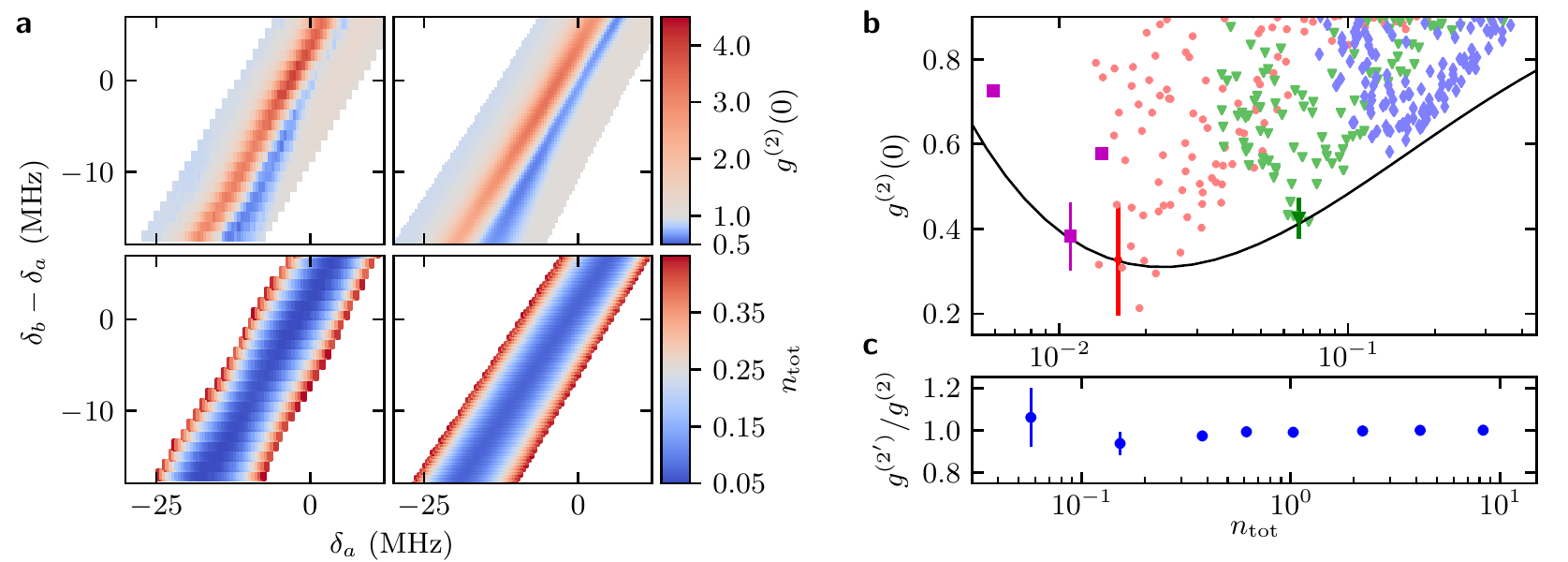}
\caption{a) Evolution of $\gd(0)$ (top row) and $\ntot$ (bottom row) as a function of the detuning $\delta_a$ and $\delta_b-\delta_a$. The left column shows experimental data that we compare to numerical simulations in the right column. The experimental pump power is -98 dBm and is adjusted in the simulation to reproduce the observed photon number. b) Evolution of $\gd(0)$ as a function of the resonator population. The light blue diamonds correspond to the dataset shown in a). The light red circles and green triangles correspond to two similar datasets measured at -104 and -101\,dBm pump power respectively. The dark red and green points with 1$\sigma$ error bars correspond to the average of light red and green data points close to the minimum of $\gd(0)$. The magenta squares show measurements at fixed detunings, which should minimize $\gd(0)$ at low power, for three different powers (-107, -105, -104\,dBm) and high statistics. The solid line is the prediction of the minimal $\gd(0)$ for our system \cite{Note1} as a function of $\ntot$. c) Experimental validation of the Gaussian state assumption. We have measured for fixed detunings all the moments of $X$ and $Y$ up to the fourth order for different pump strengths. We compute the second order correlation function with and without the Gaussian assumption and plot their ratio.}
\end{figure}

\twocolumngrid
~\\

An important figure of merit for a single photon source is the evolution of $\gd(0)$ as a function of the source brightness, which is equal to $\gamma_2 \ntot$ where $\ntot=|\alpha|^2+n$ is the resonator population. In order to minimize $\gd(0)$ for a given population, the pump strength, the pump frequency and the resonator detuning must be optimized. Experimentally, we fix the pump strength and measure $\gd(0)$ and $\ntot$ varying both $\omega_p$ and $\omega_b$ as shown in figure 4a for one pump strength. By plotting the same data points as a function of $\ntot$, we obtain a cloud of points whose lower envelope gives the minimal $\gd(0)$ as a function of $\ntot$ for our system (see figure 4b). The solid line shows the predicted minimum. Its value decreases with $\ntot$ and reaches a minimum when $\ntot$ becomes of the order of $\nth$ and then increases again when the thermal population dominates. 

In figure 4b, we also show points with error bars that are obtained by averaging over a large number of measurements close to the minimum of $\gd(0)$ at a given pump power in order to obtain a better estimate of its value. We obtain $0.38\pm0.08$, $0.33\pm0.13$ and $0.43\pm0.05$, respectively for the magenta, red, and green points. For these points, we now estimate the effect of a miscalibration in $n_h$ and $\nth$ on the value of $\gd(0)$. Assuming an error of a factor two for both quantities, $\gd(0)$ varies between 0.18 and 0.76 for the magenta point, 0.17 and 0.63 for the red point and between 0.39 and 0.49 for the green point. The magenta point is very sensitive to a change in $\nth$ because a large fraction of the resonator population is thermal. With increasing $\ntot$ and smaller thermal fraction, the systematic error decreases. 

Finally, we have checked the validity of the Gaussian assumption by measuring for a few points the moments of $X$ and $Y$ up to the fourth order. We then compute $g^{(2')}(0)=\langle a^\dagger a^\dagger a a\rangle/\langle a^\dagger a\rangle^2$ and compare it to the value of $g^{(2)}(0)$ deduced from (\ref{eq.g2Gaussian}) as shown in figure 4c. Given the statistical error bars, the ratio $g^{(2')}(0)/g^{(2)}(0)$ is consistent with one in the considered range of $\ntot$. Simulations confirm that the Gaussian hypothesis is more and more valid with decreasing $\ntot$ and we therefore expect the Gaussian assumption to be valid in the full range used in the experiment \cite{Note1}. 


In conclusion, we have observed the main features of the UPB using two coupled superconducting resonators. We found a minimal value of $\gd(0) \approx 0.4$ which is limited by the thermal population in the cavity. An intriguing question is the extension of the UPB to a large number of coupled weakly nonlinear resonators and its role in the dynamics of coherently pumped lattices of superconducting resonators \cite{Houck:2012iq}. 

\begin{acknowledgments}
The authors would like to thank Rapha\"el Weil and Sylvie Gautier for their help in the micro-fabrication of the sample. This work was partially funded by the "Investissements d'Avenir" LabEx PALM (ANR-10-LABX-0039-PALM). 
\end{acknowledgments}

\emph{Note added in proof.} -- Recently, we have become aware of a similar work in the optical domain \cite{Snijders:2018vw}.

\bibliography{biblio}

\end{document}


\title{Observation of the unconventional photon blockade effect in the microwave domain: Supplemental Material}

\author{Cyril Vaneph}
\author{Alexis Morvan}
\author{Gianluca Aiello}
\author{Mathieu F\'echant}
\author{Marco Aprili}
\author{Julien Gabelli}
\author{J\'er\^ome Est\`eve}
\affiliation{Laboratoire de Physique des Solides, CNRS, Universit\'e Paris-Sud, Universit\'e Paris-Saclay, Orsay, France}


\maketitle

\section{Master equation description}
From a numerical simulation of the microwave response of the sample, we obtain the input-output relations that relate the outgoing field $d_j^{\rm out}$ on port $j$ to the incoming field $d_j^{\rm in}$
\begin{equation}
d_j^{\rm out} = d_j^{\rm in} +\frac{\omega_0}{\sqrt{2}} ( B_{1j} a + B_{2j} b),
\label{eq.inout}  
\end{equation} 
where $B$ is a $2\times4$ matrix that describes the coupling between the two resonator modes $a$ and $b$ and the port modes. In order to derive these equations from the classical equations of motion for the field amplitudes, we have made the two usual approximations that are the rotating wave approximation and the frequency independent coupling approximation. We have checked that the coefficients of $B$ indeed vary very little on the frequency range of interest. At $\omega_0 = \pit 5.878$\,GHz, we obtain
\begin{equation}
B = 10^{-3} \times \begin{pmatrix}
14.2 & -52.0 &   0.8 &  3.9 \\
-0.8 & -3.4 & -14.2 & 54.0 \\
\end{pmatrix}.
\end{equation}
Equation (\ref{eq.inout}) can be rewritten in the more usual form
\begin{equation}
d_j^{\rm out} = d_j^{\rm in} + \sqrt{\gamma_j} c_j
\label{eq.inout_simplified}  
\end{equation}
by defining the rates $\gamma_j = (\omega_0/2)(B^2_{1j}+B^2_{2j})$ and the operator $c_j = \alpha_j a + \beta_j b$, where $\alpha_j = B_{1j}/\sqrt{B^2_{1j}+B^2_{2j}}$ and $\beta_j = B_{2j}/\sqrt{B^2_{1j}+B^2_{2j}}$. The probability per unit time for a photon to leave the system through port $j$ is $\gamma_j\langle c^\dagger_j c_j\rangle$. The master equation describing the evolution of the density matrix in our system can be written
\begin{subequations}
\label{eq.fullmeq}
\begin{eqnarray}
\frac{d \rho}{dt} & = & -\frac{i}{\hbar}\left[H, \rho\right] + \sum_{j=1}^4 \frac{\gamma_j}{2} \mathcal{D}(c_j, n_{{\rm th},j}) \rho  
+ \frac{\gamma_a}{2} \mathcal{D}(a, n_{\rm th,box})\rho + \frac{\gamma_b}{2} \mathcal{D}(b, n_{\rm th,box})\rho\\
H/\hbar & = & -\delta_a a^\dagger a - \delta_b b^\dagger b + J (a^\dagger b + b^\dagger a) - U b^\dagger b^\dagger bb + \eta_a (a + a^\dagger) + \eta_b (b + b^\dagger)\\
\mathcal{D}(c, n_{\rm th}) \rho & = &  (n_{\rm th}+1) (2c\rho c^\dagger - c^\dagger c \rho - \rho c^\dagger c) + n_{\rm th} (2c^\dagger \rho c - c c^\dagger \rho - \rho c c^\dagger). 
\end{eqnarray}
\end{subequations}
The Hamiltonian is written in the frame rotating at the pump frequency $\omega_p$ and we define the two detunings $\delta_a=\omega_p-\omega_a$ and $\delta_b=\omega_p-\omega_b$. Compared to the Hamiltonian given in the main text, we now include a pumping term for each mode. The pumping rates are given by $\eta_a = \sum_j \sqrt{\gamma_j} \alpha_{j} \langle d_j^{\rm in} \rangle$ and $\eta_b = \sum_j \sqrt{\gamma_j} \beta_{j} \langle d_j^{\rm in} \rangle$. In addition to photon loss through the ports, we also consider an intrinsic loss channel for each resonator with rates $\gamma_a$ and $\gamma_b$. Finally, we suppose that a thermal population $n_{{\rm th},j}$ with $j=1,2,3,4$ can be associated to each port. And we suppose that the intrinsic loss channels are modes of the box enclosing the sample with thermal population $n_{\rm th,box}$.

\subsection{Simplified master equation}
We can simplify the master equation if we assume that mode $a$ is only coupled to ports 1 and 2, while mode $b$ is only coupled to ports 3 and 4. This corresponds to setting to zero the four smallest elements of the $B$ matrix. The jump operators simplify to $c_1=c_2=a$, $c_3=c_4=b$ and the master equation becomes
\begin{eqnarray}
\frac{d \rho}{dt} & = & -\frac{i}{\hbar}\left[H, \rho\right] + \frac{\kappa_a}{2} \mathcal{D}(a, n_{\rm th,a})\rho + \frac{\kappa_b}{2} \mathcal{D}(b, n_{\rm th,b})\rho .
\end{eqnarray}
The total loss rates and the thermal populations are given by
\begin{subequations}
\begin{eqnarray}
\kappa_a & = & \gamma_1 + \gamma_2 + \gamma_a \\
\kappa_b & = & \gamma_3 + \gamma_4 + \gamma_b \\
n_{\rm th,a} & = & (\gamma_1 n_{\rm th,1} + \gamma_2 n_{\rm th,2} + \gamma_a n_{\rm th,box})/\kappa_a \\
n_{\rm th,b} & = & (\gamma_3 n_{\rm th,3} + \gamma_4 n_{\rm th,4} + \gamma_b n_{\rm th,box})/\kappa_b .
\end{eqnarray}
\end{subequations}
The pumping term also simplifies and, because we only pump the system through port 1 in the experiment, we obtain $\eta_b=0$. This simplified form is the one that was first studied by Liew and Savona to discover the unconventional blockade effect.    

\subsection{Numerical parameters}
Assuming the simplified form of the $B$ matrix, we obtain $\gamma_1=\pit 0.59$\,MHz, $\gamma_2=\pit 7.95$\,MHz, $\gamma_3=\pit 0.59$\,MHz and $\gamma_4=\pit 8.57$\,MHz. As explained in the text, we deduce $J=\pit 25.1$\,MHz, $U=\pit 0.25$\,MHz, $\kappa_a=\pit 10.35$\,MHz and $\kappa_b = \pit 7$\,MHz from spectroscopy measurements. We thus define $\gamma_a=\pit 1.81$\,MHz to account for the internal loss of the resonator $a$. The measured value of $\kappa_b$ is smaller than the coupling loss predicted from our microwave simulation. But, the observed resonance is slightly asymmetric, which seems to indicate a defect on the lines coupled to ports 3 and 4. This may explain the discrepancy between the simulation and the measurement. We therefore assume that $\gamma_b \approx 0$ and that $\kappa_b$ is in between $\pit 7$\,MHz and $\gamma_3 + \gamma_4 = \pit 9.2$\,MHz. 

The last parameters entering the master equation are the thermal populations $n_{\rm th,a}$ and $n_{\rm th,b}$ that we calculate as follows. The thermal occupation of the modes close to $\omega_0$ in a cable after an attenuator anchored at temperature $T$ is $D n+(1-D)n_{\rm BE}(\omega_0,T)$ where $D$ is the power attenuation factor of the attenuator, $n$ the thermal occupation before the attenuator and $n_{\rm BE}$ the Bose-Einstein distribution. For port 1, we derive the thermal populations starting from the cable outside the cryostat at a temperature of 300\,K. For port 2, we assume that the amplifier emits a black-body radiation with an effective temperature given by the amplifier noise temperature, which is 2\,K, and that each circulator behaves as a -20\,dB attenuator. We obtain $n_{\rm th,1} = 1.5 \times 10^{-2}$, $n_{\rm th,2} = 6.5 \times 10^{-4}$ and $n_{\rm th,3} \approx n_{\rm th,4} \approx n_{\rm th,box} \approx 0$. The thermal populations $n_{\rm th,a}$ and $n_{\rm th,b}$ entering the simplified master equation are the weighted averages of the port populations with weights $\gamma_i$. We finally obtain $n_{\rm th,a}=1.4 \times 10^{-3}$ and $n_{\rm th,b} \approx 0$.  

\section{Solution of the master equation}
\subsection{Pump off: Thermal state}
\begin{figure}[h!]
\includegraphics[scale=0.8]{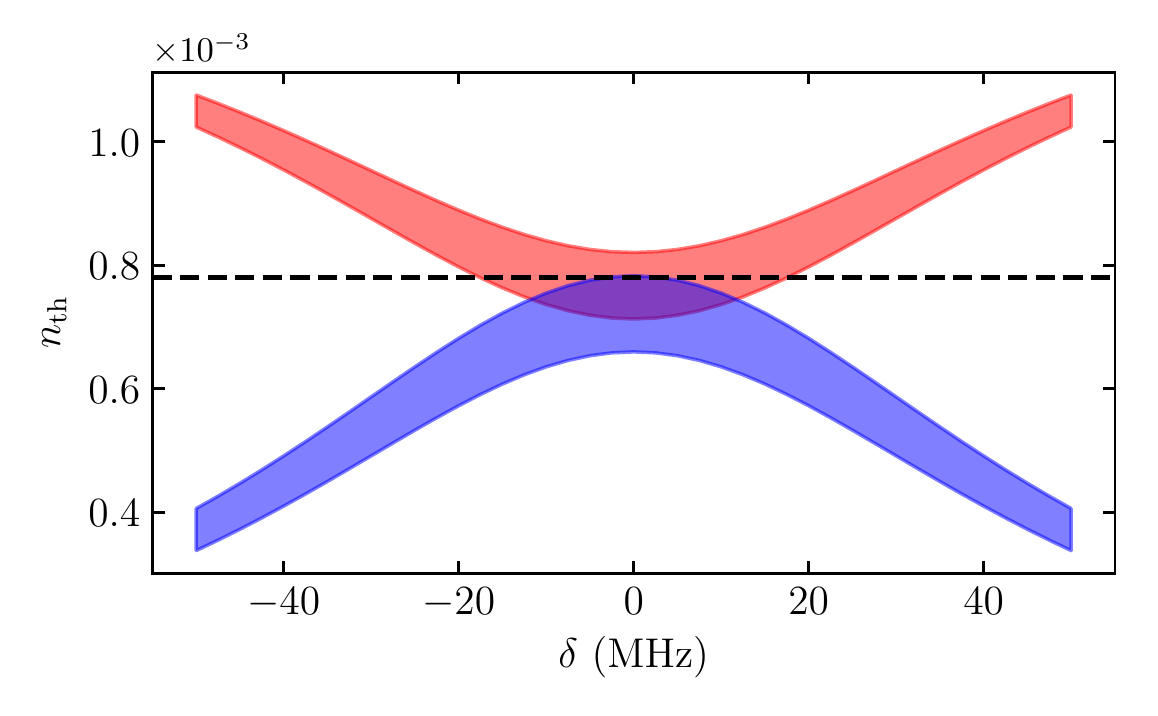}
\caption{Evolution of the thermal population of mode $a$ in red and mode $b$ in blue with the frequency of the $b$ mode. The shaded area corresponds to the prediction of the master equation for $\kappa_b/2\pi$ ranging from 7 to 10\,MHz. The horizontal line is the value that we use for the population of mode $a$ in our data analysis.\label{fig.nth}}
\end{figure}

The populations $n_{\rm th,a}$ and $n_{\rm th,b}$ correspond to the thermal populations of the $a$ and $b$ modes when the two resonators are far detuned. The population offset $\nth$ that must be added to the measurement is the population of the mode $a$ when the two resonators are close to resonance. An estimate of $\nth$ can be obtained by assuming $\kappa_a=\kappa_b=\kappa$ and $U=0$, the simplified master equation is then easily solvable and we obtain
\begin{equation}
\label{eq.nth}
\nth = \frac{\delta ^2+\kappa ^2+2 J^2}{\delta ^2+\kappa ^2+4 J^2} n_{\rm th,a} + \frac{2 J^2}{\delta ^2+\kappa ^2+4 J^2} n_{\rm th,b}, 
\end{equation}
where $\delta = \omega_b - \omega_a$. This expression leads to $\nth=7.3 \times 10^{-4}$ when $\delta=0$ assuming $\kappa=\pit 10$\,MHz. In order to obtain a more precise value, we solve the master equation as shown in figure \ref{fig.nth} for values of $\kappa_b/2\pi$ between 7 and 10\,MHz. For the analysis of all the data presented in this paper, we use $\nth = 7.8 \times 10^{-4}$ as shown by the horizontal dashed line.  

\subsection{Pump on: Displaced master equation}
\begin{figure}[h!]
\includegraphics[scale=0.8]{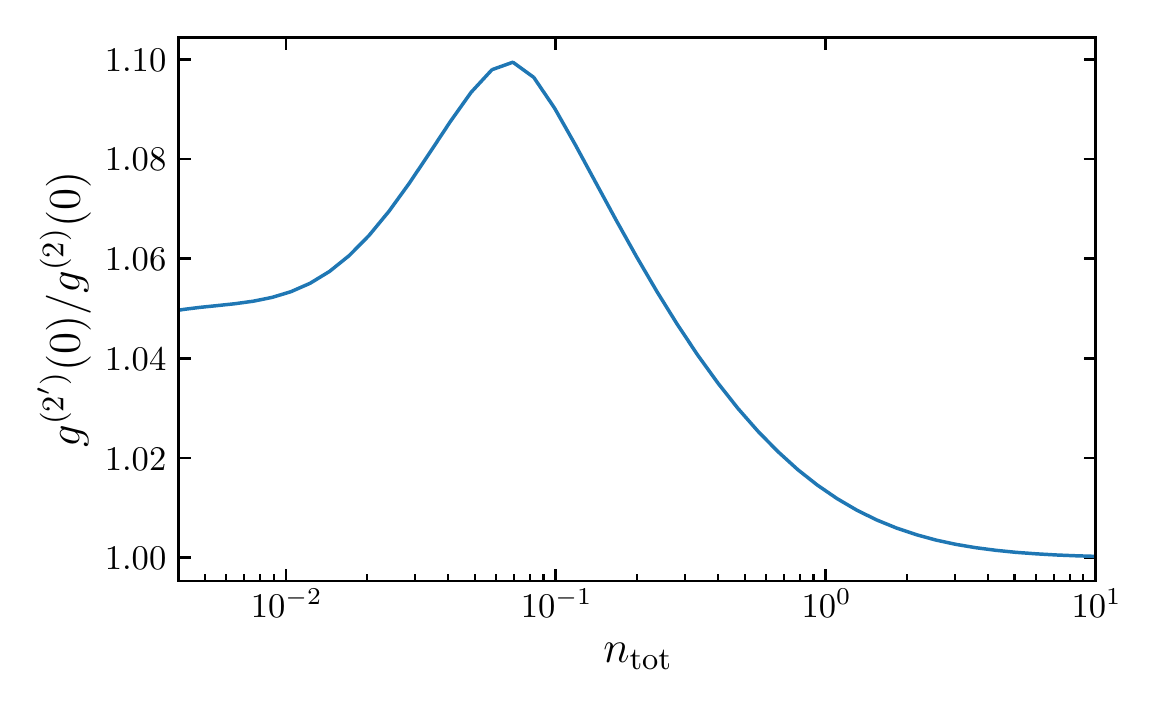}
\caption{Numerical check of the validity of the gaussian assumption. The master equation is solved for the parameters that minimize the value of $\gd(0)$ as a function of $\ntot$. We then compute $\gd(0)$ and $g^{(2')}(0)$ and plot their ratio.\label{fig.compg2}}
\end{figure}

In order to solve the master equation when the pump is on, we first solve the classical equations of motion for the mean values $\alpha = \langle a \rangle$ and $\beta = \langle b \rangle$ and obtain their steady state values. We then solve the master equation for the displaced density matrix $D^\dagger \rho D$ where $D=D(\alpha)\otimes D(\beta)$ and $D(\alpha)$ is the displacement operator. The Hilbert space is truncated to the lowest four Fock states for each mode. We then calculate the different observables
\begin{subequations} 
\begin{eqnarray}
\ntot & = & |\alpha|^2 + \langle a^\dagger a \rangle \\
g^{(2')}(0) & = & \frac{\langle a^\dagger a^\dagger a a \rangle  + \alpha^2 \langle a^\dagger a^\dagger\rangle +
(\alpha^*)^2 \langle a a\rangle + 4 |\alpha|^2 \langle a^\dagger a\rangle + |\alpha|^4 }{\ntot^2} \\
\gd(0) & = & \frac{\langle a^\dagger  a^\dagger \rangle \langle  aa \rangle +2\langle a^\dagger  a \rangle^2  + \alpha^2 \langle a^\dagger a^\dagger\rangle +
(\alpha^*)^2 \langle a a\rangle + 4 |\alpha|^2 \langle a^\dagger a\rangle + |\alpha|^4 }{\ntot^2},
\end{eqnarray}
\end{subequations}
where the average is taken for the density matrix that is solution of the displaced master equation. The expression $\gd(0)$ is only valid when the state is gaussian and is equivalent to equation (2) of the main text. In figure \ref{fig.compg2}, we plot the ratio $g^{(2')}(0)/\gd(0)$ as in the figure 4c of the main text. Here, the detunings are chosen to minimize $\gd(0)$ for a given $\ntot$. In the regime $\ntot \approx 3 \times 10^{-2}$ where $\gd(0)$ is minimal, the error is about 7\,\% which is well below our statistical and systematic error. 

The simulations shown in the figures 2 and 4 of the main text correspond to the solution of the simplified master equation using the method explained above and extracting $\gd(0)$ using equation (8c). For the value of $\kappa_b$, we use $\pit 7$\,MHz, which leads to a better agreement with the measured $\gd(0)$ than higher values of $\kappa_b$. In figure 2, we suppose $\delta_b=\delta_a$ and we adjust the value of $\eta$ to reproduce the measured displacement $|\alpha|^2$. The curve of figure 4b is obtained through a numerical minimization of $\gd(0)$ over $\delta_a$ and $\delta_b$ for different values of $\eta$. 




\section{Experimental determination of the moments}
Starting from the input-output relation (\ref{eq.inout_simplified}) and making the assumption that the coupling from the $b$ mode to the port 2 is negligible, the field entering the mixer is given by
\begin{equation}
a_{\rm ampl} = \sqrt{G_{\rm ampl}} \left[  \sqrt{G_{\rm att}}(\sqrt{\gamma_2} a + d_2^{\rm in}) +  \sqrt{G_{\rm att}-1} h^\dagger_{\rm att} \right] +  \sqrt{G_{\rm ampl}-1} \, h^\dagger_{\rm ampl},
\end{equation}
where $G_{\rm ampl}$ is the amplification gain, $G_{\rm att}$ the attenuation between the sample and the amplifier, and $h_{\rm att}$, $h_{\rm ampl}$, $d_2^{\rm in}$ are bosonic fields that we suppose to be in thermal states. Because the effective temperatures of $h_{\rm att}$ and $d_2^{\rm in}$ are much smaller than the one $T_{\rm ampl}=2$\,K of the amplifier noise and assuming $G_{\rm ampl} \gg 1$, we obtain 
\begin{subequations}
\begin{align}
a_{\rm ampl} & = \sqrt{G}(a + h^\dagger) \\
h & \approx  \frac{1}{\sqrt{\gamma_2}  \sqrt{G_{\rm att}} } h_{\rm ampl},
\end{align}
\end{subequations}
where $G$ is a global gain factor and $h$ the effective thermal noise impeding the measurement.

\subsection{Correction of the IQ mixer imperfection}
An important step in the data analysis procedure is the correction of the IQ mixer imperfection. As explained in the main text, our goal is to obtain the DC components $\bar{X}(t)$, $\bar{Y}(t)$ and the AC components $X(t)$, $Y(t)$ of the quadratures of the field $a + h^\dagger$. We denote by $\bar{X}_r(t)$, $\bar{Y}_r(t)$, $X_r(t)$, $Y_r(t)$ the raw data obtained by filtering and digitizing the two outputs of the IQ mixer. They are related to the actual quadratures through the following relations in the frequency domain
\begin{subequations}
\label{truemodel}
\begin{align}
\bar{X}_r[0] &= \sqrt{G} \, g_{X}^{\rm DC}  X[0] \\
\bar{Y}_r[0] &= \sqrt{G} \, g_{Y}^{\rm DC} (Y[0] + \epsilon   X[0]) \\
X_{r}[\omega] &= \sqrt{G}  \, g_{X}[\omega]  X[\omega] \\
Y_{r}[\omega] &= \sqrt{G}  \, g_{Y}[\omega] (Y[\omega] + \epsilon   X[\omega] ), 
\end{align}
\end{subequations}
where $\epsilon$ is the quadrature phase deviation of the mixer, $g_{X}[\omega]$, $g_{Y}[\omega]$ are the transfer functions of the AC acquisition chains, and $g_{X}^{\rm DC}$, $g_{Y}^{\rm DC}$ are DC gains. From  the digitized $X_r(t)$ and $Y_r(t)$ traces, we compute the moments $\langle X_r^i Y_r^j \rangle$, which can be expressed in the frequency domain as
\begin{equation}
\label{eq.def_moments}
\langle X_r^i Y_r^j \rangle = \frac{1}{(2 \pi)^{i+j}} \int_{-\infty}^\infty \langle X_{r}[\omega_1] \ldots X_{r}[\omega_i] Y_{r}[\omega_{i+1}] \ldots Y_r[\omega_{i+j}]\rangle  \, \delta(\omega_1 + \ldots + \omega_{i+j}) \, d\omega_1 \ldots d\omega_{i+j}.
\end{equation}
In order to extract the moments of $X$ and $Y$, we rely on the fact that the two transfer functions $g_X[\omega]$ and $g_Y[\omega]$ have a very similar frequency dependence. We therefore assume $\sqrt{G} g_X[\omega] \approx \sqrt{G_X} g[\omega]$ and $\sqrt{G} g_Y[\omega] \approx \sqrt{G_Y} g[\omega]$. Equations (\ref{truemodel}c,d) then become
\begin{subequations}
\label{model}
\begin{align}
X_{r}[\omega] &= \sqrt{G_X}  \, g[\omega]  X[\omega] \\
Y_{r}[\omega] &= \sqrt{G_Y}  \, g[\omega] (Y[\omega] + \epsilon   X[\omega] ). 
\end{align}
\end{subequations}
We can rewrite (\ref{eq.def_moments}) as 
\begin{equation}
\langle X_r^i Y_r^j \rangle = G_X^{i/2} G_Y^{j/2} \langle X^i (Y + \epsilon X)^j \rangle,
\label{eq.raw_moments}
\end{equation}
where the moments on the rhs of the equation are defined as
\begin{equation}
\langle X^i Y^j \rangle = \frac{1}{(2 \pi)^{i+j}} \int_{-\infty}^\infty g[\omega_1] \ldots g[\omega_{i+j}] \langle X[\omega_1]  \ldots X[\omega_i] Y[\omega_{i+1}] \ldots Y[\omega_{i+j}]\rangle  \, \delta(\omega_1 + \ldots + \omega_{i+j}) \, d\omega_1 \ldots d\omega_{i+j}.
\label{eq.true_moments}
\end{equation}
The transfer functions of the filters must be chosen such that $g[\omega]\approx 1$ over the frequency range where $X[\omega]$ and $Y[\omega]$ are non zero. Equation (\ref{eq.raw_moments}) can be solved to obtain the moments of $X$ and $Y$ from the moments of $X_r$ and $Y_r$. For example, the second order moments are given by
\begin{subequations}
\label{eq.moments2}
\begin{align}
\mo{X^{2}} &= \frac{\mo{X_{r}^{2}}}{G_{X}}   \\
\mo{XY} &= \frac{\mo{X_{r}Y_{r}}}{\sqrt{G_{X}G_{Y}}} - \epsilon \frac{\mo{X_{r}^{2}}}{G_{X}}   \\
\mo{Y^{2}} &= \frac{\mo{Y_{r}^{2}}}{G_{Y}} - 2\epsilon \frac{\mo{X_{r}Y_{r}}}{\sqrt{G_{X}G_{Y}}} + \epsilon^{2}  \frac{\mo{X_{r}^{2}}}{G_{X}}.  
\end{align}
\end{subequations}
The expressions of the higher order moments are given in Appendix A. Finally, from an independent calibration of the gain of the DC chains, we rescale the measured DC components such that equations (\ref{truemodel}a,b) can be rewritten as
\begin{subequations}
\begin{align}
\bar{X}_r[0] &= \sqrt{G_X}  X[0] \\
\bar{Y}_r[0] &= \sqrt{G_Y} (Y[0] + \epsilon  X[0]).
\end{align}
\end{subequations}

\subsection{Calibration of the gains and phase deviation}
In order to obtain $G_X$, $G_Y$ and $\epsilon$, we consider the measured moments when the pump is off, in which case the field arriving on the mixer is the thermal field $h$. Equation (\ref{eq.true_moments}) gives 
\begin{align}
\mo{X^{2}}_0 & = \frac{1}{2 \pi} \int_{-\infty}^\infty |g[\omega]|^2 \langle X[\omega] X[-\omega] \rangle \, d\omega= n_h  \nonumber  \\
\mo{XY}_0    & = \frac{1}{2 \pi} \int_{-\infty}^\infty |g[\omega]|^2 \langle X[\omega] Y[-\omega] \rangle \, d\omega= 0    \nonumber  \\
\mo{Y^{2}}_0 & = \frac{1}{2 \pi} \int_{-\infty}^\infty |g[\omega]|^2 \langle Y[\omega] Y[-\omega] \rangle \, d\omega= n_h  \nonumber  
\end{align}
with $n_h = k_B T_{\rm ampl} \Delta f /( G_{\rm att} \gamma_2 \hbar \omega_0)$ and $\Delta f=\int_{-\infty}^\infty |g[\omega]|^2 \, d\omega/2\pi = 24$\,MHz. From equations (\ref{eq.moments2}), we obtain
\begin{align}
G_{X} &= \frac{\mo{X_{r}^{2}}_{0}}{n_h} \nonumber \\
G_{Y} &= \frac{\mo{Y_{r}^{2}}_{0} \mo{X_{r}^{2}}_{0} - \mo{X_{r}Y_{r}}_{0}^{2} }{n_h \mo{X_{r}^{2}}_{0}}  \\
\epsilon &=  \frac{\mo{X_{r}Y_{r}}_{0}}{ n_h  \sqrt{G_{X}G_{Y} }}. \nonumber
\end{align}
Using these expressions and the expressions of Appendix A, we obtain the moments of $X$ and $Y$ from the measured moments of $X_r$ and $Y_r$. The estimation of $G_X$, $G_Y$ and $\epsilon$ is repeated for every measurement point in order to take into account experimental drifts.   

\subsection{Statistical error}
Data are acquired by packets of $8.4\times10^7$ points. The moments are estimated from a pump on and a pump off packet as explained above. We then average the moments over $N_p$ packets before computing $\gd(0)$ using equation (2) of the main text. Because this equation is non-linear, if $N_p$ is too small the probability distribution of $\gd(0)$ for different measurements is not gaussian. We use $N_p>20$ to avoid this problem.  

\section{Appendix A}
We use the following expressions to obtain the moments of $X$ and $Y$ as a function of the experimentally measured moments of $X_r$ and $Y_r$:
\begin{align}
\mo{\bar{X}} &= \frac{\mo{\bar{X}_{r}}}{\sqrt{G_{X}}} \nonumber  \\
\mo{\bar{Y}} &= \frac{\mo{\bar{Y}_{r}}}{\sqrt{G_{Y}}} -  \epsilon \frac{\mo{\bar{X}_{r}}}{\sqrt{G_{X}}} \nonumber  \\
\mo{X^{2}} &= \frac{\mo{X_{r}^{2}}}{G_{X}}  \nonumber  \\
\mo{XY} &= \frac{\mo{X_{r}Y_{r}}}{\sqrt{G_{X}G_{Y}}} - \epsilon \frac{\mo{X_{r}^{2}}}{G_{X}}  \nonumber  \\
\mo{Y^{2}} &= \frac{\mo{Y_{r}^{2}}}{G_{Y}} - 2\epsilon \frac{\mo{X_{r}Y_{r}}}{\sqrt{G_{X}G_{Y}}} + \epsilon^{2}  \frac{\mo{X_{r}^{2}}}{G_{X}}  \nonumber  \\
\mo{X^{3}} &= \frac{\mo{X_{r}^{3}}}{G_{X}^{3/2}}  \nonumber  \\
\mo{X^{2}Y} &= -\frac{\epsilon  \sqrt{G_Y} \mo{X_{r}^3}-\sqrt{G_X} \mo{X_{r}^2 Y_{r}}}{G_X^{3/2} \sqrt{G_Y}} \nonumber\\
\mo{XY^{2}} &= -\frac{-\epsilon ^2 G_Y \mo{X_{r}^3}+2 \epsilon  \sqrt{G_X} \sqrt{G_Y} \mo{X_{r}^2 Y_{r}}-G_X \mo{X_{r}Y_{r}^2}}{G_X^{3/2} G_Y}\nonumber \\
\mo{Y^{3}} &=-\frac{\epsilon ^3 G_Y^{3/2} \mo{X_{r}^3}-3 \epsilon ^2 \sqrt{G_X} G_Y \mo{X_{r}^2 Y_{r}}+3 \epsilon  G_X \sqrt{G_Y} \mo{X_{r} Y_{r}^2}-G_X^{3/2} \mo{Y_{r}^3}}{G_X^{3/2} G_Y^{3/2}} \nonumber \\
\mo{X^{4}} &= \frac{\mo{X_{r}^{4}}}{G_{X}^{2}} \nonumber \\
\mo{X^{3}Y} &=-\frac{\epsilon  \sqrt{G_Y} \mo{X_{r}^4}-\sqrt{G_X} \mo{X_{r}^3 Y_{r}}}{G_X^2 \sqrt{G_Y}} \nonumber \\
\mo{X^{2}Y^{2}} &= -\frac{-\epsilon^2 G_Y \mo{X_{r}^4}+2 \epsilon  \sqrt{G_X} \sqrt{G_Y} \mo{X_{r}^3 Y_{r}} - G_X \mo{X_{r}^2 Y_{r}^2}}{G_X^2 G_Y} \nonumber \\
\mo{XY^{3}} &=-\frac{\epsilon ^3 G_Y^{3/2} \mo{X_{r}^4}-3 \epsilon ^2 \sqrt{G_X} G_Y \mo{X_{r}^3 Y_{r}}+3 \epsilon  G_X  \sqrt{G_Y} \mo{X_{r}^2 Y_{r}^2}-G_X^{3/2} \mo{X_{r} Y_{r}^3}}{G_X^2 G_Y^{3/2}}\nonumber \\
\mo{Y^{4}} &= -\frac{-\epsilon ^4 G_Y^2 \mo{X_{r}^4}+4 \epsilon ^3 \sqrt{G_X} G_Y^{3/2} \mo{X_{r}^3 Y_{r}}-6 \epsilon^2 G_X G_Y \mo{X_{r}^2 Y_{r}^2}}{G_X^2 G_Y^2} \nonumber \\
& +\frac{4 \epsilon  G_X^{3/2} \sqrt{G_Y} \mo{X_{r} Y_{r}^3}-G_X^2 \mo{Y_{r}^4}}{G_X^2 G_Y^2}\nonumber 
\end{align}